\documentclass[12pt]{article}
\usepackage{epsfig}
\usepackage{floatflt} 
\usepackage{caption}
\usepackage{graphicx}

\def\beq{\begin{equation}}
\def\eeq#1{\label{#1}\end{equation}}
\def\eeqn{\end{equation}}

\def\beqa{\begin{eqnarray}}
\def\eeqa#1{\label{#1}\end{eqnarray}}
\def\eeqan{\end{eqnarray}}

\let\bar=\overbar

\def\Dslash{\not{\hbox{\kern-4pt $D$}}}
\def\dslash{\not{\hbox{\kern-2pt $\del$}}}

\def\msb{{\bar{\ssstyle M \kern -1pt S}}}

\def\Title#1{\begin{center} {\Large {\bf #1} } \end{center}}

\begin{document}

\Title{Heavy-Flavour Production in Pb--Pb collisions at the LHC with ALICE}

\bigskip\bigskip


\begin{raggedright}  

{\it Yvonne Pachmayer\index{Pachmayer, Y.} for the ALICE Collaboration\\
Physikalisches Institut der Universit\"at Heidelberg, Im Neuenheimer Feld 226\\
D-69120 Heidelberg, Germany}
\bigskip\bigskip
\end{raggedright}

\noindent Results on open heavy-flavour production in \mbox{Pb--Pb} collisions at $\sqrt{s_{\rm NN}}$ = 2.76 TeV measured with ALICE at the LHC are presented. The nuclear modification factors, extracted in three different channels, show a strong suppression in central collisions. The measured D-meson azimuthal anisotropy indicates a non-zero $v_{2}$, which is similar to the one of charged hadrons. 

\section{Introduction}
According to calculations of lattice Quantum Chromodynamics (QCD) \cite{lQCD} matter is expected to undergo a phase transition from a hadronic phase to a Quark-Gluon Plasma (QGP) at energy densities larger than 0.5 $\rm GeV/fm^{3}$. In the QGP phase quarks and gluons are deconfined and chiral symmetry is restored. Collisions of Pb nuclei at LHC generate an extended volume of high energy density (initial energy density $\approx$ 15 $\rm GeV/fm^{3}$). Heavy quarks, i.e. charm and beauty quarks, are among the most interesting and powerful probes to investigate the properties of the QGP. They are, due to their heavy mass, produced on a very short time scale in initial hard scattering processes and thus they experience the whole history of the collision. They interact strongly with the deconfined medium, loose energy and may participate in the collective expansion. Therefore they enable us to study parton energy loss as well as its color and quark mass dependence. The measurement of the elliptic flow $v_{2}$ of D and B mesons probes on the one hand the degree of thermalization of massive quarks in the medium at low transverse momentum and on the other hand the path length dependence of energy loss at high transverse momentum. \newline
\noindent ALICE (A Large Ion Collider Experiment) is well suited to detect and identify open charm and beauty hadrons due to a momentum resolution better than 2\% for \mbox{$p_{\rm T} < $ 20 GeV/$c$,} a transverse impact parameter resolution better than 65(20) $\rm \mu$m for a $p_{\rm T} >$  1(20) GeV/$c$ and because of various systems for particle identification, e.g. Time Projection Chamber (TPC), Time of Flight system (TOF), Muon Spectrometer. The experiment and its heavy-quark detection performance are described in \mbox{detail in \cite{ALICEexperiment}.} 
In this paper open heavy-flavour production in Pb--Pb collisions is presented in the following channels:
\begin{itemize}
\item Open charm and beauty - reconstruction of electrons from semi-electronic decays:
\noindent D, B $\rm \rightarrow e + X$ in $\rm |\textit{y}_{e}| < 0.8$ \newline 
The electrons were identified using the signals in the TOF and the TPC. To extract electrons from heavy-flavour hadron decays a data-tuned Monte Carlo cocktail of electrons from background sources was subtracted from the inclusive electron spectrum. Further details can be found in \cite{HFEPbPb}.
\vspace{-0.3cm}
\item Open charm and beauty - reconstruction of muons from semi-muonic decays: \newline
\noindent D, B $\rm \rightarrow \mu + X$ in $\rm -4 < \textit{y}_{\mu} < -2.5$ \newline 
Single muons were measured in the Muon Spectrometer by matching reconstructed tracks with tracks in the muon trigger chambers \cite{ALICEexperiment}. Background muons from the decay-in-flight of light hadrons were estimated by extrapolating $\rm K^{\pm}$ and $\rm \pi^{\pm}$ spectra measured at mid-rapidity to forward rapidities and by applying the corresponding decay kinematics. Further information can be found in \cite{MuonPbPb}.
\vspace{-0.3cm}
\item Open charm - fully reconstructed hadronic decays: \newline
\noindent $\rm D^{0} \rightarrow K^{-} \pi^{+}$, $\rm D^{+} \rightarrow K^{-} \pi^{+}\pi^{+}$, $\rm D^{*+} \rightarrow D^{0} \pi^{+}$ and charge conjugates in $\rm |\textit{y}| < 0.5$ \newline
\noindent The reconstuction is based on the invariant mass analysis of fully reconstructed decay topologies displaced with respect to the primary vertex. The large combinatorial background was reduced by identifying charged pions and kaons in the TPC and TOF. The correction for feed-down from B-meson decays was done using FONLL calculations \cite{FONLLcalc}. More details on the analysis are described in \cite{DPbPb}.
\end{itemize}

\noindent The results presented in this contribution were obtained from the first two Pb--Pb runs at \mbox{$\sqrt{s_{\rm NN}}=2.76$ TeV}, which took place in 2010 and 2011. The Silicon Pixel Detector (SPD) at mid-rapidity and the forward VZERO scintillator counters provide a minimum-bias (MB) interaction trigger, and are also used to derive the centrality of the collisions. In total 17M MB Pb--Pb collisions (2010) were used for analysis. The elliptic flow results are based on 9.5M events (2011) in the 30-50\% centrality class.

\section{Open heavy-flavour suppression}

\begin{figure*}[htb]
     \begin{minipage}{0.43\textwidth}
      \centering
       \includegraphics[width=1\textwidth]{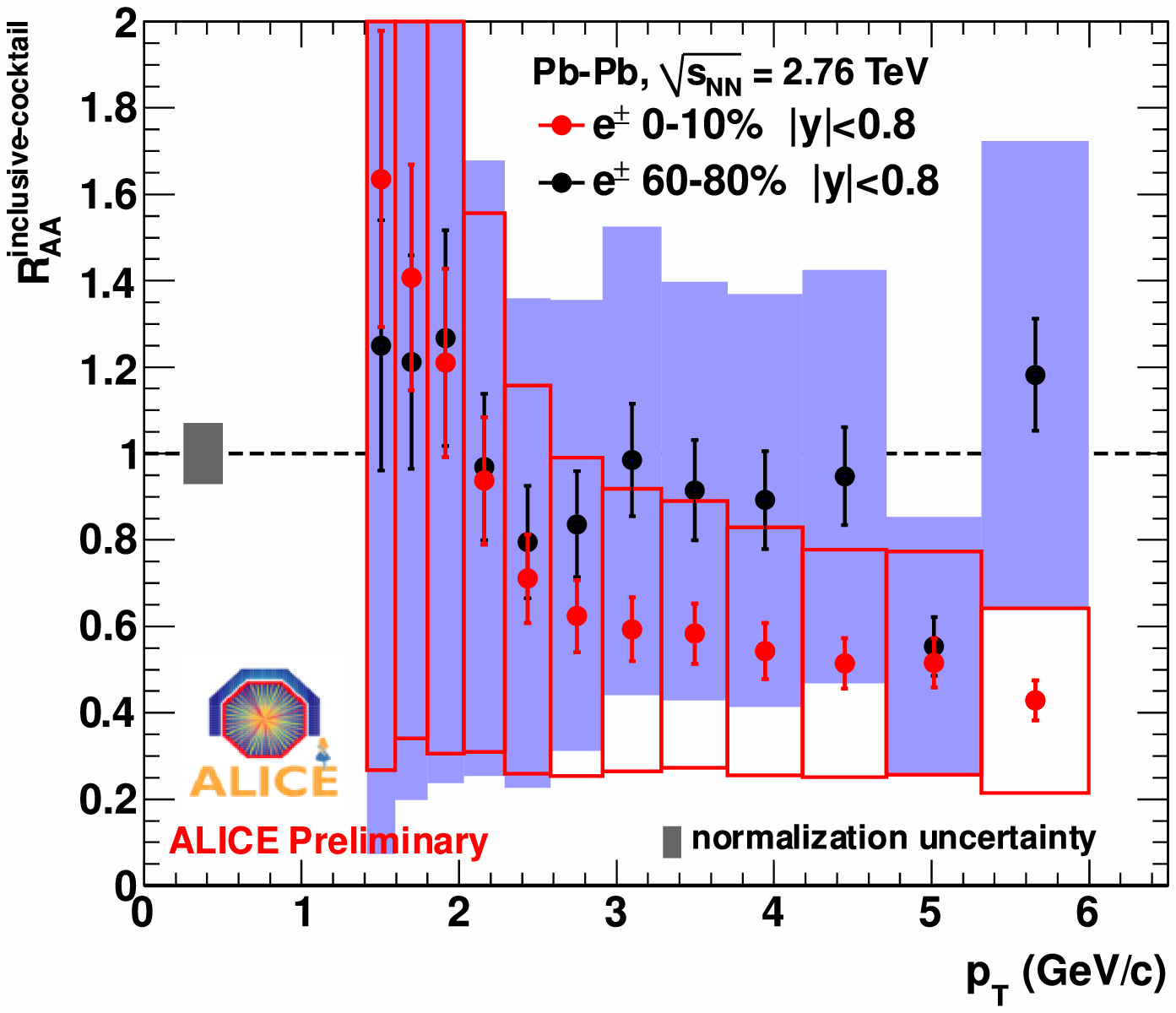}
       \caption{\small $R_{\rm AA}$ of background-subtracted electrons for central and peripheral \mbox{Pb--Pb} collisions.}\label{figure1}
     \end{minipage}
     \hspace{1.6cm}
     \begin{minipage}{0.4\textwidth}
      \centering
    \includegraphics[width=0.9\textwidth]{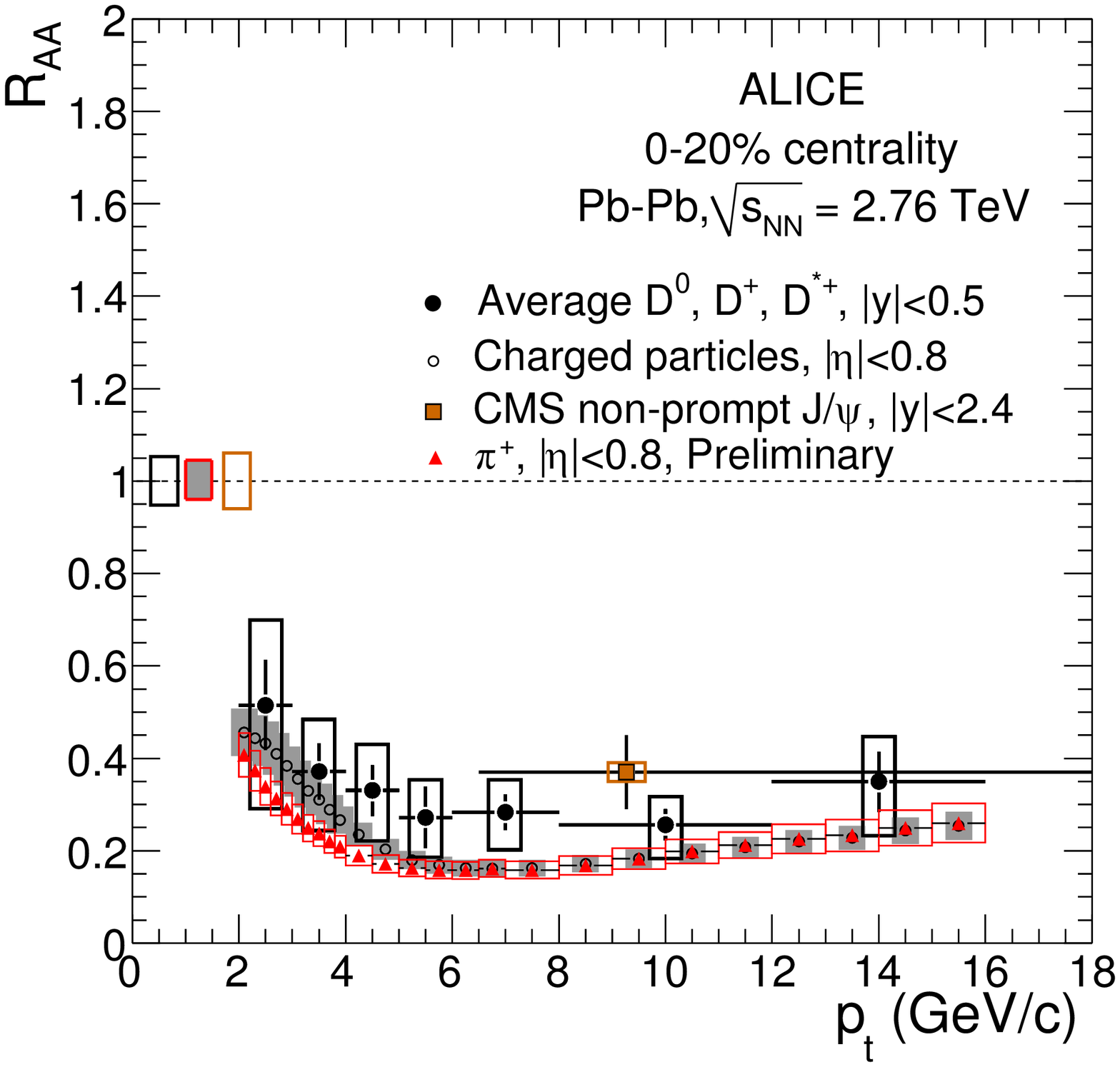} 
       \caption{\small Comparison of the $R_{\rm AA}$ of D mesons, charged hadrons and pions as well as non-prompt J/$\psi$ mesons in the 0-20\% centrality class.}\label{figure2}
     \end{minipage}
   \end{figure*}

An observable to quantify the interaction of hard partons with the medium is the nuclear modification factor $R_{\rm AA}$, where one compares particle production in Pb--Pb collisions with pp collisions at the same centre-of-mass energy:
$R_{\rm AA}(p_{\rm T}) = \frac{1}{ \langle T_{\rm AA} \rangle} \cdot \frac{dN_{\rm AA}/dp_{\rm T}}{d\sigma_{\rm pp}/dp_{\rm T}}$. $\langle T_{\rm AA} \rangle$ denotes the average nuclear overlap function for a given centrality range, $dN_{\rm AA}/dp_{\rm T}$ and $d\sigma_{\rm pp}/dp_{\rm T}$ represent the particle yield in nucleus-nucleus collisions and the cross section in pp collisions, respectively. \newline
\noindent \mbox{Figure \ref{figure1}} shows the nuclear modification factor of background-subtracted electrons for the centrality ranges 0-10\% and 60-80\%. The pp reference spectrum is obtained by scaling the measured spectrum of electrons from heavy-flavour decays in pp at \mbox{$\sqrt{s}$ = 7 TeV} to 2.76 TeV based on FONLL calculations \cite{ExtrapolationPaper}. In contrast to peripheral events, a suppression of a factor of \mbox{1.5-4} is found for 0-10\% central collisions in the $p_{\rm T}$ region 3.5-6 GeV/$c$, where charm and beauty decays a priori dominate \cite{HFEPbPb}. Including the Transition Radiation Detector (TRD) and the Electromagnetic Calorimeter for particle identification will lead to an extension of the $p_{\rm T}$ range of the $R_{\rm AA}$ towards low and high $p_{\rm T}$ as well as to a reduction of the systematic uncertainty.
In the near future, the charm and beauty contributions will be disentangled via secondary vertexing. \newline
\noindent Also, the heavy-flavour decay muon $R_{\rm AA}$ for $p_{\rm T}$ = 4-10 GeV/$c$ yields a suppression of a factor of 3-4 in central collisions (0-10\%) with no significant $p_{\rm T}$ dependence \cite{MuonPbPb}. The FONLL \cite{FONLLcalc} prediction indicates that beauty-decay muons dominate for \mbox{$p_{\rm T} >$ 6 GeV/$c$.} \newline
\noindent The average $R_{\rm AA}$ of three D-meson species ($\rm D^{0}, D^{+}$ and $\rm D^{*+}$) is shown in Fig. \ref{figure2}. The pp reference at $\sqrt{s}$ = 2.76 TeV was obtained by scaling the measured 7 TeV spectrum with FONLL calculations \cite{ExtrapolationPaper}. The respective spectrum was cross-checked against the measured result of a short pp run taken at $\sqrt{s}$ = 2.76 TeV. For the 0-20\% centrality class a suppression of a factor of 3-4 for $p_{\rm T} >$ 5 GeV/$c$ is found. The supression is reduced when going to more peripheral collisions and at lower transverse momentum. Soon the errors will be reduced by including data from the second Pb--Pb run (2011), where 6-7 times more statistics in the 0-7.5\% centrality range were collected. Using next-to-leading order (NLO) pQCD calculations, the effect of shadowing on the D-meson $R_{\rm AA}$ was estimated to be $\sim$15\% for \mbox{$p_{\rm T} >$ 6 GeV/$c$.} Thus the visible strong suppression is most likely a final state effect. The upcoming p-Pb run (scheduled for 2013) will allow to measure directly the initial state effects. 
\vspace{-0.4cm}
\vspace{-0.1cm}
\noindent In Fig. \ref{figure2} the $R_{\rm AA}$ of prompt D mesons is compared with the one of charged hadrons, pions \cite{ChargedPion} and non-prompt J/$\psi$ from B decays \cite{CMSJPSI}. There is an indication for $R_{\rm AA}^{\rm D} > R_{\rm AA}^{\pi}$, whereas the suppression for non-prompt J/$\psi$ seems weaker. However a more differential and precise measurement of the $p_{\rm T}$ dependence is necessary for a conclusive statement on color and mass ordering effects.

\section{Elliptic flow $v_{2}$}

\begin{floatingfigure}[l]{0.45\textwidth}
 \vspace{-0.6cm}
\begin{center}
\includegraphics[width=0.45\textwidth]{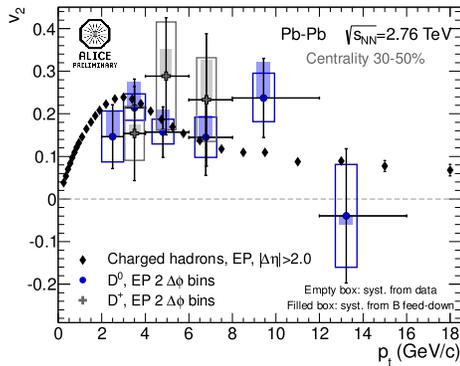}\end{center}\vspace{-0.4cm}\captionsetup{singlelinecheck=off,justification=RaggedRight}
\vspace{-0.5cm}\caption{\small $\rm D^{0}$, $\rm D^{+}$ and charged hadron $v_{2}$ in the 30-50\% centrality class.}\label{figure3}
\end{floatingfigure}  
\noindent In non-central heavy ion collisions the spatial anisotropy with respect to the reaction plane (defined by the beam axis and the impact parameter of the colliding nuclei) is translated into a momentum anisotropy due to multiple collisions.
The magnitude of this asymmetry can be quantified using a Fourier decomposition of the $p_{\rm T}$-dependent azimuthal distribution of particles w.r.t. the estimated reaction plane (called event plane). The second harmonic is called elliptic flow coefficient, $v_{2}$. \newline
\noindent The measurements of the $\rm D^{0}$ and $\rm D^{+}$ $v_{2}$ (see Fig. \ref{figure3}) indicate a non-zero $v_{2}$ in the $p_{\rm T}$ range 2-6 GeV/$c$. These results are similar to the charged hadron $v_{2}$ measured with ALICE in the same rapidity region. \newline

\noindent In conclusion, we have measured in several decay channels a strong suppression of heavy-flavour production in central Pb--Pb collisions at $\sqrt{s_{\rm NN}}$ = 2.76 TeV. There is a hint of a lower suppression for D mesons than for pions. The measured elliptic flow of D mesons seems non-zero and is within uncertainties comparable with the one of charged hadrons. These results indicate strong coupling of heavy quarks to the medium.
In the near future, the contributions of charm and beauty quarks will be separated where applicable. The $p_{\rm T}$ range will be extended as well as uncertainties reduced by increasing statistics and improving particle identification. Finally, initial and final state effects will be disentangled by measuring p-Pb collisions scheduled for the beginning of 2013.


\begin{thebibliography}{99}

\vspace{-0.3cm}
\bibitem{lQCD}
A. Bazavov et al., Phys. Rev. D85 (2012) 054503.
\vspace{-0.3cm}
\bibitem{ALICEexperiment}
 B.~Alessandro et al. [ALICE Collaboration], J. Phys. G30 (2004) 1517, 32 (2006) 1295 and J. Instrum. 3 (2008) S08002.
\vspace{-0.3cm}
\bibitem{HFEPbPb}
Y.~Pachmayer for the ALICE Collaboration, J. Phys. G38 (2011) 124186.
\vspace{-0.3cm}
\bibitem{MuonPbPb}
B.~Abelev et al. [ALICE Collaboration], arXiv:1205.6443 [hep-ex] (2012).
\vspace{-0.3cm}
\bibitem{FONLLcalc}
 M.~Cacciari, M.~Greco and P.~Nason, JHEP 9805 (1998) 007.
\vspace{-0.3cm}
\bibitem{DPbPb}
B.~Abelev et al. [ALICE Collaboration], arXiv:1203.2160 [nucl-ex](2012).
\vspace{-0.3cm}
\bibitem{ExtrapolationPaper}
R.~Averbeck et al., arXiv:1107.3243 [hep-ph] (2011).
\vspace{-0.3cm}
\bibitem{CMSJPSI}
CMS Collaboration, arXiv:1201.5069 [nucl-ex] (2012).
\vspace{-0.3cm}
\bibitem{ChargedPion}
H.~Appelsh\"auser for the ALICE Collaboration, J. Phys. G38 (2011) 124014.



\end{thebibliography}
\end{document}